\documentclass[onecolumn]{aastex62}
\usepackage[T1]{fontenc}
\usepackage[utf8]{inputenc}
\usepackage{amsmath}
\usepackage{textcomp}
\usepackage[normalem]{ulem}
\usepackage{etoolbox}

\makeatletter
\let\@left\left
\let\@right\right
\renewcommand{\left}{\mathopen{}\mathclose\bgroup\@left}
\renewcommand{\right}{\aftergroup\egroup\@right}
\makeatother
\newcommand{\paren}[1]{\left( #1 \right)}
\newcommand{\sbrack}[1]{\left[ #1 \right]}
\newcommand{\given}{\,\middle\vert\,}
\newcommand{\of}{\paren}
\newcommand{\px}{\pi}
\newcommand{\gaia}{\emph{Gaia}}
\newcommand{\masyr}{\ensuremath{{\mathrm{mas}\,\mathrm{yr}^{-1}}}}

\received{June 14, 2018}
\revised{June 29, 2018}
\submitjournal{ApJ}

\shorttitle{Binary pulsar distances}
\shortauthors{Jennings et al.}

\begin{document}

\title{Binary Pulsar Distances and Velocities from \gaia\ Data Release 2}

\correspondingauthor{Ross Jennings}
\email{rjj58@cornell.edu}

\author[0000-0003-1082-2342]{Ross J. Jennings}
\affiliation{Department of Astronomy, Cornell University, Ithaca, NY 14853, USA}
\author[0000-0001-6295-2881]{David L. Kaplan}
\affiliation{Center for Gravitation, Cosmology and Astrophysics, Department of Physics, University of Wisconsin--Milwaukee, P.O. Box 413, Milwaukee, WI 53201, USA}
\author[0000-0002-2878-1502]{Shami Chatterjee}
\affiliation{Department of Astronomy and Cornell Center for Astrophysics and Planetary Science, Cornell University, Ithaca, NY 14853, USA}
\author[0000-0002-4049-1882]{James M. Cordes}
\affiliation{Department of Astronomy and Cornell Center for Astrophysics and Planetary Science, Cornell University, Ithaca, NY 14853, USA}
\author[0000-0001-9434-3837]{Adam T. Deller}
\affiliation{Centre for Astrophysics and Supercomputing, Swinburne University of Technology, PO Box 218, Hawthorn, VIC 3122, Australia}
\begin{abstract}
The second data release from the \gaia\ mission (\gaia\ DR2) includes, among its billion entries, astrometric parameters for  binary companions to a number of known pulsars, including white dwarf companions to millisecond pulsars and the non-degenerate components of so-called ``black widow'' and ``redback'' systems. 
We find 22 such counterparts in DR2, of which 12 have statistically significant measurements of parallax. 
These DR2 optical proper motions and parallaxes provide new measurements of the distances and transverse velocities of the associated pulsars.
For the most part, the results agree with existing radio interferometric and pulsar timing-based astrometry, as well as other distance estimates based on photometry or associations, and for some pulsars they provide the best known distance and velocity estimates. 
In particular, two of these pulsars have no previous distance measurement: PSR~J1227$-$4853, for which \gaia\ measures a parallax of $0.62\pm 0.16$ mas, and PSR~J1431$-$4715, with a \gaia\ parallax of $0.64\pm 0.16$ mas.   
Using the \gaia\ distance measurements, we find that dispersion measure-based distance estimates calculated using the NE2001 and YMW16 Galactic electron density models are on average slightly underestimated, which may be a selection effect due to the over-representation of pulsars at high Galactic latitudes in the present \gaia\ sample. 
While the \gaia\ DR2 results do not quite match the precision that can be achieved by dedicated pulsar timing or radio interferometry, taken together they constitute a small but important improvement to the pulsar distance scale, and the subset of millisecond pulsars with distances measured by \gaia\ may help improve the sensitivity of pulsar timing arrays to nanohertz gravitational waves. 
\end{abstract}

\section{Introduction}

The estimation of distances is a fundamental problem in astronomy, and since observers measure apparent quantities, distance estimates underpin much of astrophysics as well. For neutron stars, distance measurements (whether direct or indirect) enable a range of inferences, from constraints on the nuclear equation of state using radius measurements \citep[e.g.,][]{kvka02,hh09,shs+18,of16}, to the physics of energy transport and the conversion of Poynting flux to particle flows in neutron star winds \citep[e.g.,][]{aaa+13,ska+16}.
Precision astrometry of neutron stars  allows the determination of birth sites and associations \citep[e.g.,][]{vcc04,kvka07,kcga08,tnhm10,kvc+15} and high transverse velocity measurements constrain the physics of supernova core collapse \citep[e.g.,][]{cvb+05} as well as the evolution of close binary systems.  Distances and transverse velocities are also important to determining  both the underlying spin-down rates of pulsars  and the intrinsic orbital decay rates of binaries, correcting for the kinematic Shklovskii effect~\citep{shklovskii} and Galactic acceleration \citep[e.g.,][]{acw99,wkk00}, which are important for the accurate determination of spin-down luminosities \citep[e.g.,][]{gsl+16} and relativistic orbital decay \citep[e.g.,][]{dt91}.
However, distances to neutron stars are difficult to measure, and for most radio pulsars, they are only estimated indirectly through the observed pulse dispersion measure (DM $\equiv \int_0^d n_e\,ds$, the line-of-sight integral of the electron density). These DM-based distance estimates rely on models of the Galactic electron density distribution such as that of \citet[][henceforth NE2001]{cl02} or that of \citet[][henceforth YMW16]{ymw17}. Precise, independent distance measurements are therefore vital, both in their own right and for their use in calibrating such models and thereby improving distance estimates for the rest of the radio pulsar population.

Independent distance estimates for radio pulsars come from a variety of sources, including limits based on \ion{H}{1} absorption or associations with stellar clusters, spectroscopic parallaxes of binary companions, and other techniques, but the primary techniques rely on geometric parallax, which is equivalent to measuring the curvature of the pulse wavefront over the Earth's orbit. Parallax can be measured via annual variations in pulse times of arrival (requiring timing precision $\lesssim$1~$\mu$s for distances of $\sim$1~kpc; e.g., \citealt{tbm+99,lk12,rhc+16,dcl+16,gsl+16,mnf+16}) or from astrometric imaging observations with sub-milliarcsecond precision, which can be accomplished with Very Long Baseline Interferometry (VLBI; e.g., \citealt{cbv+09,dvk+16}).
Both techniques (pulsar timing and interferometry) yield astrometry tied to the International Celestial Reference Frame (ICRF; \citealt{fgj+15}), as defined by extragalactic radio quasars \citep{mae+98}. However, pulsar timing measurements depend on the position of the earth relative to the Solar System barycenter, and so can be affected by errors in the Solar System ephemeris~\citep{fglm84,wch+17}. Interferometric measurements use a geocentric frame and so do not suffer from that issue.

The \gaia\ spacecraft \citep{gaia}, launched in 2013, is conducting an all-sky optical survey of more than 1 billion astrophysical sources with magnitudes $\lesssim 21$. Like its predecessor {\em Hipparcos} \citep{plk+97}, \gaia\ solves simultaneously for the astrometric parameters of sources distributed across the sky. The results are connected to the ICRF through optical quasars \citep{mkl+18}. The second \gaia\ data release (DR2, based on the first two years of data of a planned five year mission) contains the first astrometric solutions based entirely on \gaia\ data, including, for the majority of the catalog, the first measurements of parallax and proper motion~\citep{gaia-dr2}.

Neutron stars are not the most obvious targets for optical astrometry, given their small sizes and lack of optical emission in all but a few cases (e.g., \citealt{m11}; discussed further below). However, the companions of pulsars in binary systems offer opportune targets.  
We present results from a search through the \gaia~DR2 catalog, where we have identified the companions to known binary pulsars by matching positions and proper motions, as well as the expected magnitude of the companions themselves where available, in order to weed out false positives.  
We describe our methods for source identification in \S~\ref{sec:methods}, present our results in \S~\ref{sec:results}, including comparisons with previous distance measurements and estimates from Galactic electron density models, and conclude with a discussion of our results and future prospects in \S~\ref{sec:discuss}. 

\section{Methods}
\label{sec:methods}
\subsection{Identification of sources}
\label{sec:id}
We identified probable companions to pulsars in the \gaia\ catalog based on their positions. Of the 2424 pulsars outside of globular clusters listed in the ATNF pulsar catalog \citep[][v.\ 1.57]{mhth05,mhtt16}, we selected the 188 that are in binary systems.  We further selected those that have position uncertainties $<1\arcsec$ in both coordinates.  This is largely to exclude  pulsars  that cannot be reliably matched against \gaia\ astrometry, and effectively selects for those that have full phase-coherent timing solutions spanning at least a year.  We selected binary pulsars because, with the exception of the Crab pulsar\footnote{The Crab pulsar is detected by \gaia, but the parallax is not detected at a statistically significant level. The proper motion of $\mu_\alpha=-11.8\pm0.2\,\masyr$, $\mu_\delta=+2.6\pm0.2\,\masyr$ is consistent with the proper motion measured using archival {\em Hubble Space Telescope} observations by \citet[][$\mu_\alpha=-11.8\pm0.4\pm0.5\,\masyr$, $\mu_\delta=+4.4\pm0.4\pm0.5\,\masyr$]{kcga08}. }, no pulsar is intrinsically bright enough at optical wavelengths to be visible with \gaia\ \citep[e.g.,][]{m11}.

The resulting 155 sources were then cross-matched with \gaia\ DR2.  Our cross-match used a radius of $1\arcsec$ after correcting the position of the pulsar to the epoch of \gaia\ (J2015.5) using the measured proper motion from the ATNF catalog.  In most cases this is still far larger than the position uncertainty, with the majority of genuine counterparts agreeing to $<0\farcs1$. In principle there could be even better agreement between pulsar timing positions and \gaia, but discrepancies between timing positions and the ICRF to which both \gaia\ and VLBI should be tied are present at the level of a few mas \citep{wch+17}.  Moreover, there are covariances between astrometric and other parameters in pulsar binary fitting that make some proper motions less reliable than their formal uncertainties would indicate, and some pulsar positions are only known from optical observations (notably that of PSR~J1417$-$4402; see below).  
Therefore we examined all sources with potential matches out to $1\arcsec$ by hand, consulting the literature for cases with known counterparts and examining images from the Pan-STARRS1 3$\pi$ survey \citep{cmm+16} for sources north of $\delta=-30\degr$ and  the SkyMapper Southern Sky Survey \citep{wol+18} for sources south of $\delta=-30\degr$ where there were ambiguous cases.

In almost every case, the potential association of a pulsar with a \gaia\ source was obvious and supported by previous detections in the literature (including requiring the \gaia\ match to have a similar brightness to that reported in the literature).
However, in six cases---those of PSRs~J1056$-$7117, J1125$-$6014, J1435$-$6100, J1543$-$5149, J1755$-$3716, and B1953+29---the putative matches appear from visual inspection to be unrelated foreground stars, inconsistent with the expected masses of the companions and distances to the systems \citep[e.g.,][]{mcp+14,btb+15}.  For J1431$-$4715, whose white dwarf companion had not previously been detected (\citealt{btb+15} predict $V>23$), we have verified that the companion is correct through spectroscopy (Kaiser et al., in prep.). For B1957+20 visual inspection showed that the \gaia\ source was an unrelated star $0\farcs7$ away from the pulsar, despite the pulsar having a variable companion with a V-band magnitude which reaches a minimum of about 20 \citep{kdf88,cvpr95}.
In all cases we did a further test and estimated a rate of false detection based on the number of \gaia\ sources within $1\arcmin$ with magnitudes brighter than the putative companion.  We found typical false positive rates of $10^{-5}$, and as high as $10^{-3}$ for a few sources where either the proper motion was very high or the initial position measurement was several decades ago.

Only in the cases of PSRs J0045$-$7319, J1417$-$4402,  J1723$-$2837, and J2129$-$0429 were the position offsets greater than $0\farcs2$.  However, all four of these have particularly bright companions with secure identifications.  For  PSR~J0045$-$7319 the timing solution is over 20 years old and was complicated by the significant variations in spin and orbital properties encountered in the system \citep{kbm+96}.
For PSR~J1417$-$4402 the position is obtained from optical measurements of the companion star \citep{scc+15} and lacks the typical radio pulsar precision, but we can also be sure of the identification.  For PSR~J1723$-$2837 we can again be certain of the identification of a bright companion from the literature \citep{cls+13} with radial velocity confirmation. Finally, for PSR~J2129$-$0429, even though the companion position  is offset from the pulsar position by $1\farcs3$, the companion was identified through radial velocity and photometric variations and is confirmed to be the same object, with the apparent offset between the optical position and the radio position already noted by \citet{bkb+16}.

We list all the sources with confident \gaia\ counterparts in Table~\ref{tab:other}, along with their dispersion measure distances and any other distance measurement found in the literature.  Such distances include VLBI parallaxes, timing parallaxes,  OB associations (for the Be-binary systems), spectroscopic parallaxes, and modeling of distorted companions in tight binaries.

\subsection{Parallax inversion and Lutz-Kelker correction}
\label{sec:lk}
In a field of objects with uniform density in space, more objects are located at larger distances from the observer than at smaller ones, because a spherical shell of a given thickness has a volume that scales as the square of its radius. As \citet{lk73} observed, this has important implications for estimating the distance to an object from its trigonometric parallax. In effect, they argue that in such a case the prior distribution of the distance $d$ to an object,  $p\of{d}$,  should be proportional to $d^2$. In terms of the parallax $\px$, this translates into $p\of{\px}\propto\px^{-4}$. In general, if the objects under consideration have space density $\rho\of{d,\theta,\phi}$, the prior on the distance $d$ to an object known to lie in the direction $(\theta,\phi)$ should have the form $p\of{d\given\theta,\phi}\propto d^2\rho\of{d,\theta,\phi}$.

As is typical, we assume the likelihood function for the true parallax $\px$ to have the form of a Gaussian centered on its measured value $\hat{\px}$ with uncertainty $\sigma_\px$, so that the posterior distribution for $d$ takes the form
\begin{equation}
p\of{d\given\hat{\px},\theta,\phi}\propto d^2\rho\of{d,\theta,\phi}\exp\sbrack{-\frac{(1/d-\hat{\px})^2}{2\sigma_\px^2}}.
\end{equation}
In the case of constant space density, as originally observed by Lutz \& Kelker, this has the unfortunate consequence that the posterior distribution $p\of{d\given{\px}}$ for $d$ diverges as $d\to\infty$. When the uncertainty on the parallax $\sigma_\px$ is $\ll{\px}$, the divergence occurs only for extremely large distances and can be ignored in practice. On the other hand, if $\rho\of{d,\theta,\phi}$ decreases rapidly enough as $d\to\infty$, the divergence disappears entirely.

With this in mind, we adopt a prior on the distance $d$ to each pulsar based on a model of the Galactic pulsar population. This approach is similar to that advocated by \citet{brf+18} for estimating distances from \gaia\ parallax data, but uses a pulsar-specific model for the space density $\rho$. Our model is based on that of \citet{lfl+06}, but we adopt a larger vertical scale height of 500\,pc for millisecond pulsars, following \citet{vwc12}. We use a DM-based upper bound on distance in a forthcoming analysis of the pulsar velocity distribution (Jennings et al., in prep.), but it is not included in the analysis here because its effect is insignificant. When the line of sight to the pulsar points away from the Galactic center, the effect of our prior can be to decrease the distance estimate, unlike the classical Lutz-Kelker correction based on an assumed uniform space density, which always increases the distance estimate. 

Our prior does not account for all effects which could potentially bias parallax-based distance estimates. In particular, the fact that nearer objects are more likely to be detected due to their greater apparent brightness means that our Lutz-Kelker corrected distances may be systematically too large. Furthermore, the \gaia\ parallax estimates are subject to a systematic zero point offset, which has been estimated to average $-0.029$ mas but can vary as a function of source brightness, color, and position \citep{lhb+18}. We correct for this in the only way presently possible, by adding $0.029$ mas to the central value $\hat{\px}$ of the parallax likelihood function in calculating our distance estimates. In Table~\ref{tab:measured}, in addition to giving Lutz-Kelker corrected distances $d_\mathrm{LK}$, we also give distances $d_\px$ estimated by assuming a uniform prior on parallax. In both cases, we give the mode of the posterior distribution as a point estimate of $d$, and indicate the 16th and 84th percentiles of the distribution as errors. The distances $d_\px$ are compared with previous distance estimates in Figs.~\ref{fig:compare} and \ref{fig:compare2}. The conclusions for $d_\mathrm{LK}$, however, are essentially the same. For our sample of pulsars, the Lutz-Kelker correction increases the distance estimates by no more than 30\% (in most cases less than 25\%), and in all cases the mode of the corrected distance distribution lies within the 68\% credible region of the uncorrected distribution.

\subsection{Velocity estimation}

In Table~\ref{tab:measured}, we give three estimates of the component of the velocity of each pulsar transverse to the line of sight. All three estimates are calculated based entirely on \gaia\ parallax and proper motion data. The first and second, labelled $v_{\perp,\px}$ and $v_{\perp,\mathrm{LK}}$, are estimates of the pulsar velocity with respect to the solar system barycenter, making use of the distance estimates $d_{\px}$ and $d_{\mathrm{LK}}$, respectively. The third, $v_{\perp,\mathrm{DGR}}$, is an estimate of the pulsar velocity with respect to its standard of rest, taking into account the effects of differential Galactic rotation and the Sun's peculiar velocity with respect to the local standard of rest. This estimate is also based on $d_{\mathrm{LK}}$. In all three cases, the estimates are based on the marginal distribution of the perpendicular velocity, assuming the proper motions follow normal distributions with means and standard deviations as given by \gaia\ DR2, the distance follows the appropriate posterior distribution, and the distance and proper motion are independent. As with the distance estimates, we give the mode as a point estimate and the 16th and 84th percentiles as errors. Differential Galactic rotation is calculated based on a simple model of Galactic motion in which stars follow circular orbits around the center of the galaxy with a constant velocity $\Theta_0$. To maintain consistency with the Galactic pulsar population model of \citet{lfl+06}, we set $\Theta_0=220$ km\,s$^{-1}$ and the distance $R_\sun$ from the Sun to the center of the Galaxy to $8.5$ kpc, their 1985 IAU standard values~\citep{klb86}. The components of the Sun's peculiar velocity are taken from~\citet{srb+10}.

\section{Results}
\label{sec:results}

\begin{figure}
\plotone{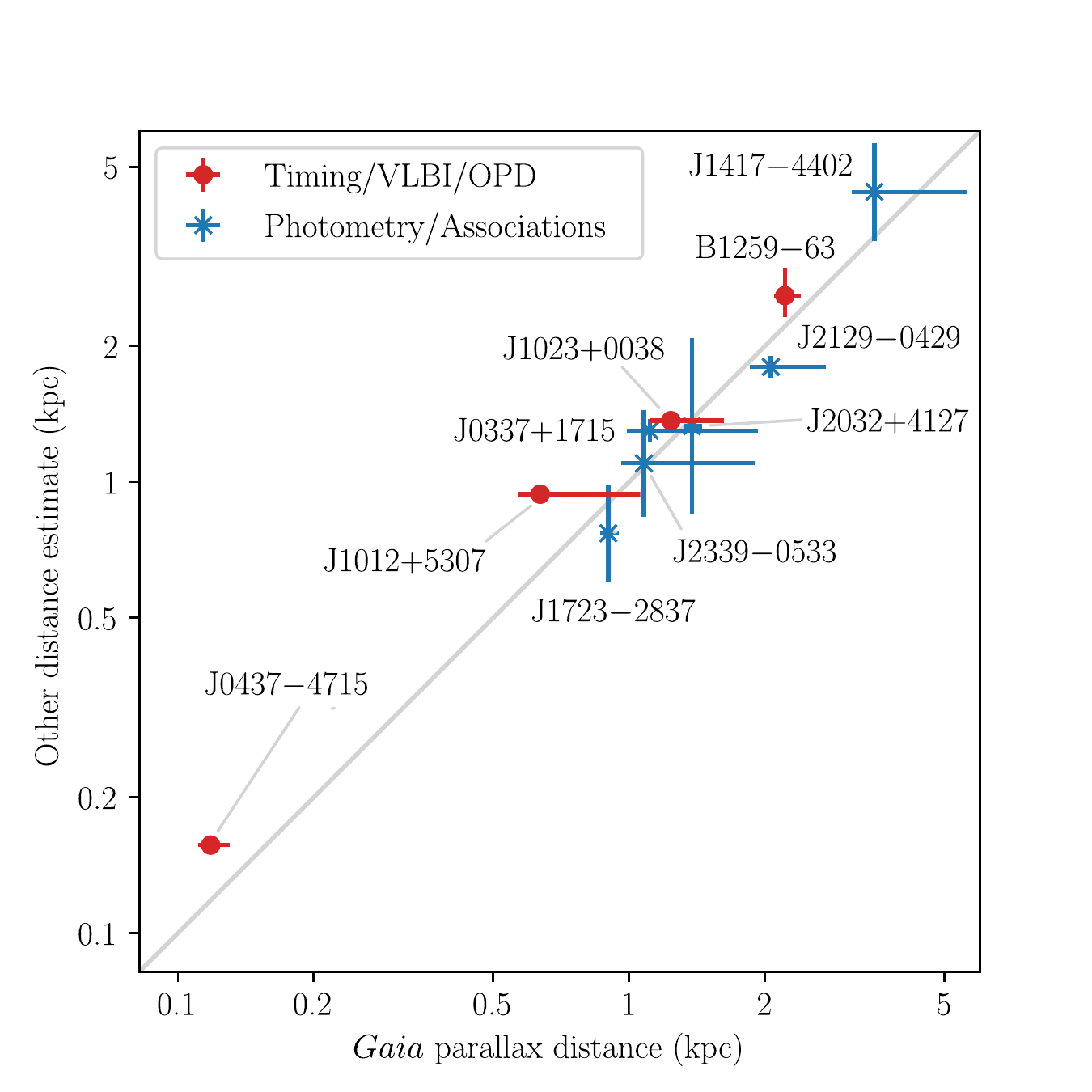}
\caption{\label{fig:compare} Comparison of \gaia\ parallax distances with the previous estimates of distance given in Table~\ref{tab:other}. Error bars represent 1-$\sigma$ uncertainties or 68\% credible intervals. For PSRs~J1417$-$4402 and J1723$-$2837, which have distance determined assuming the companion overflows its Roche lobe, 25\% error bars have been added for illustrative purposes. The \gaia\ distances  are those determined without a Lutz-Kelker correction ($d_{\px}$  in Table~\ref{tab:derived}).  
}
\end{figure}

The \gaia\ DR2 parallax and proper motion measurements are given in Table~\ref{tab:measured}. Pulsars with more significant parallax measurements (less than 40\% uncertainty in parallax) are given in the upper section of the table, and pulsars whose companions are readily identifiable in \gaia\ DR2 but which have less significant parallax measurements are given in the lower section. Previous distance estimates for the pulsars identified in Table~\ref{tab:measured} are given in Table~\ref{tab:other}, and these estimates are compared with the uncorrected  \gaia\ distances ($d_{\px}$) in Figure~\ref{fig:compare}.  Overall the agreement is good, and suggests that many of the alternate (non-astrometric) means of distance estimation for the pulsars have been reliable.

\begin{figure}
\plotone{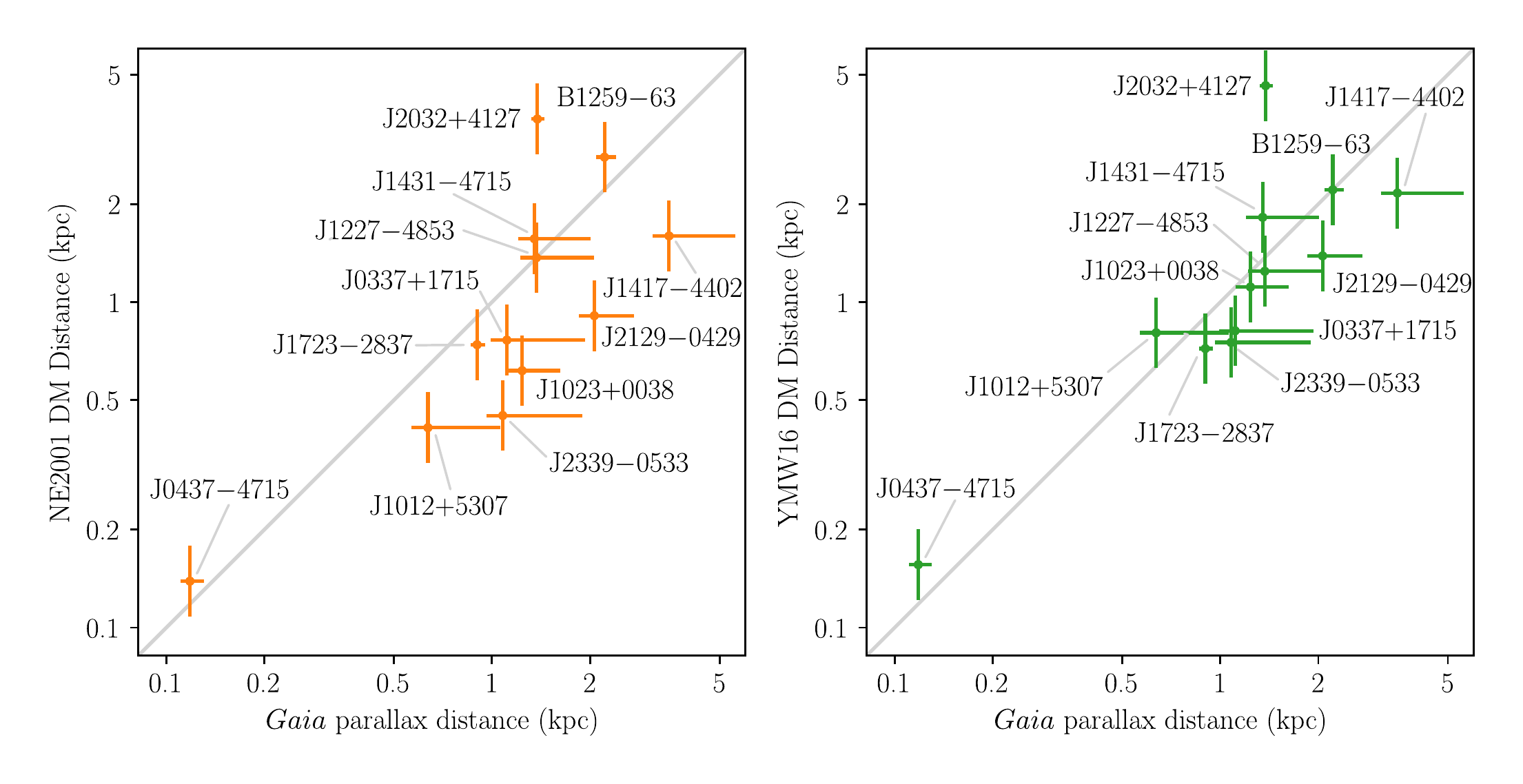}
\caption{\label{fig:compare2} 
Comparison of \gaia\ parallax distances with DM distances calculated using the NE2001 \citep[left;][]{cl02} and YMW16 \citep[right;][]{ymw17} Galactic electron density models. As in Figure~\ref{fig:compare}, the \gaia\ distances  are those determined without a Lutz-Kelker correction ($d_{\px}$  in Table~\ref{tab:derived}), and the error bars represent 1-$\sigma$ uncertainties. The error bars on the DM distances are illustrative, and are estimated at 25\%. PSR~J0045$-$7319, which is in the Small Magellanic Cloud, is not shown.\\
}
\end{figure}

\begin{figure}
\plotone{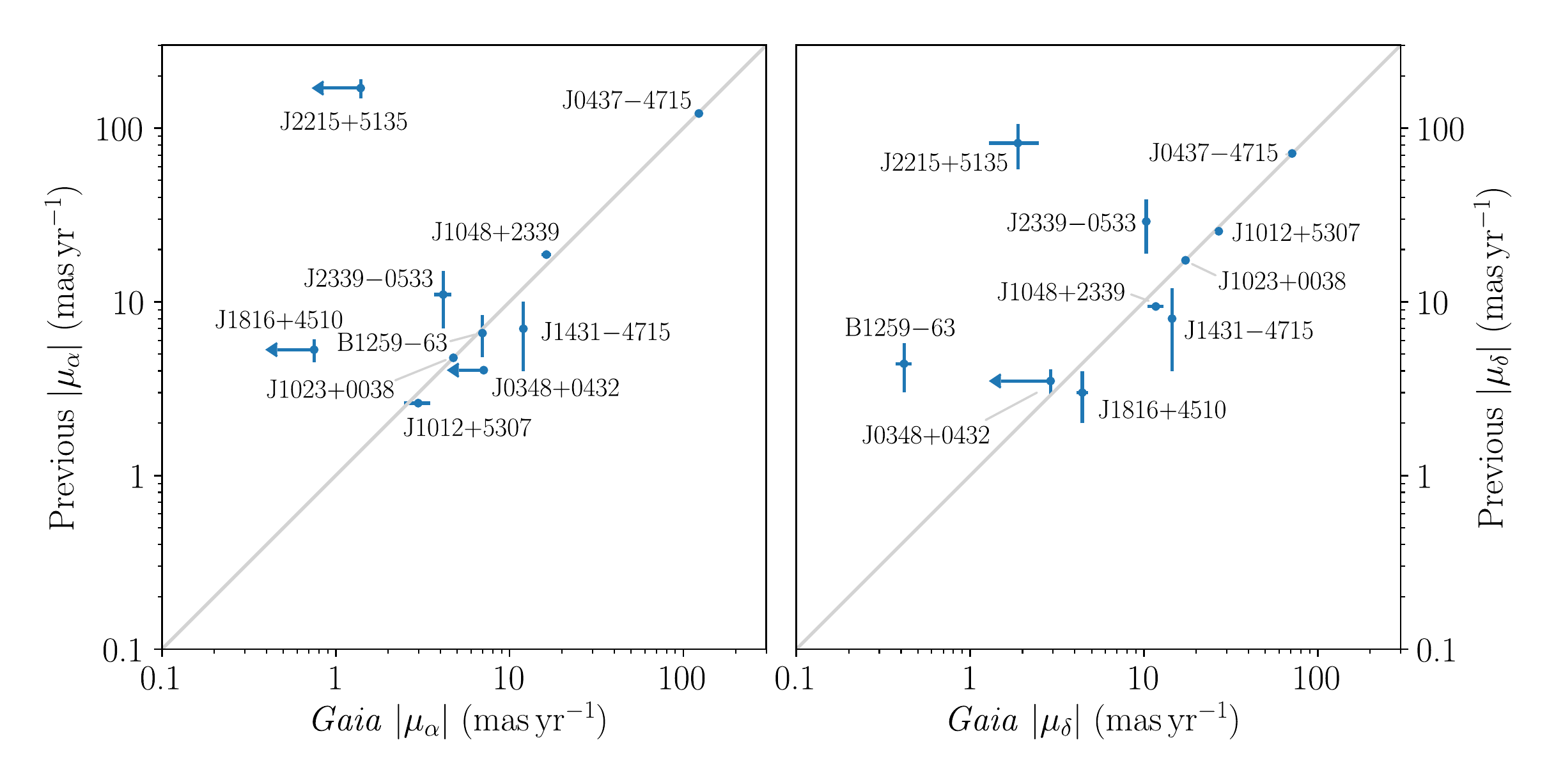}
\vspace{-2em}
\caption{Comparison of \gaia\ proper motions with previous proper motion measurements from the ATNF pulsar catalog \citep{mhth05,mhtt16}. Proper motions in right ascension $\alpha$ include the geometric correction ($\mu_\alpha=\dot{\alpha}\cos\delta$).  We plot the absolute values of the proper motions to allow for better comparison over a large dynamic range.  Sources without 3$\sigma$ detections of \gaia\ proper motions are plotted as 2$\sigma$ upper limits.
\label{fig:pm}}
\end{figure}

\begin{deluxetable}{ l l c c c c c c l l }
\tablecaption{Previously observed characteristics of pulsars with companions detected in \gaia\ DR2\label{tab:other}}
\tablehead{
\colhead{PSR} & \colhead{System Type\tablenotemark{a}} & \colhead{DM} & \multicolumn{3}{c}{Distance}
& \colhead{Other Method\tablenotemark{d}} & \colhead{Refs.}\\ \cline{4-6}
 & & & \colhead{NE2001\tablenotemark{b}} & \colhead{YMW16\tablenotemark{c}} & \colhead{Other}\\
 & & \colhead{(pc\,cm$^{-3}$)} & \colhead{(kpc)}& \colhead{(kpc)}& \colhead{(kpc)} & &
}
\decimals
\startdata
\multicolumn{5}{l}{Pulsars with significant \gaia\ parallax measurements}\\
\cline{1-5}
\object[PSR J0337+1715]{J0337+1715} & Triple & \phn$21.3$ &0.76&0.81& $1.30\pm0.08$ & Photometry & 1 \\
\object[PSR J0437-4715]{J0437$-$4715} & He WD & \phn$20.4$ &0.14&0.16& $0.15679 \pm 0.00025$ & OPD & 2\\
 & & &&& $0.1563\pm 0.0013$ & VLBI & 3 \\
\object[PSR J1012+5307]{J1012+5307} & He WD & \phn\phn$9.0$ &0.41&0.80& $0.94\pm 0.03$ & OPD & 4\\
\object[PSR J1023+0038]{J1023+0038} & Transitional & \phn$14.3$ &0.62&1.11& $1.368_{-0.039}^{+0.042}$ & VLBI & 5\\
\object[PSR J1227-4853]{J1227$-$4853} & Transitional & \phn$43.4$ &1.37&1.24& \nodata & \nodata & 6\\
\object[PSR B1259-63]{B1259$-$63} & Be Binary & $146.7$ &2.79&2.21& $2.59^{+0.37}_{-0.28}$ & VLBI & 7\\
& &  &&& $2.3\pm 0.4$ & Association & 8\\
\object[PSR J1417-4402]{J1417$-$4402} & Redback & \phn$55.0$ &1.60&2.16& 4.4 & RLO & 9,\,10\\
\object[PSR J1431-4715]{J1431$-$4715} & He WD & \phn$59.4$ &1.57&1.82& \nodata & \nodata & 11\\
\object[PSR J1723-2837]{J1723$-$2837} & Redback & \phn$19.7$ &0.74&0.72& 0.77 & RLO & 12\\
\object[PSR J2032+4127]{J2032+4127} & Be Binary & $114.7$ &3.65&4.62& $1.33\pm0.60$ & Association & 13\\
\object[PSR J2129-0429]{J2129$-$0429} & Redback & \phn$16.9$ &0.91&1.39& $1.8\pm0.1$ & Spectroscopy & 14\\
\object[PSR J2339-0533]{J2339$-$0533} & Redback & \phn\phn$8.7$&0.45&0.75& $1.1\pm0.3$ & Spectroscopy & 15,\,16\\
&  &  &&& 0.7 & Photometry & 17\\
\hline
\multicolumn{5}{l}{Pulsars with low significance \gaia\ parallax measurements}\\
\cline{1-5}
\object[PSR J0045-7319]{J0045$-$7319} & MS & $105.4$ &$>36$&58.7\tablenotemark{e}& $60\pm4\phn$ & Association (SMC) & 18\\
\object[PSR J0348+0432]{J0348+0432} & He WD & \phn$40.5$ &2.08&2.26& $2.1\pm0.2$ & Spectroscopy & 19\\
\object[PSR J1024-0719]{J1024$-$0719} & MS & \phn\phn$6.5$ &0.39 & 0.38 & $1.13\pm 0.18$ & Timing & 20\\
 & & &&& $1.08\pm0.04$ & Spectroscopy & 21,\,22\\
\object[PSR J1048+2339]{J1048+2339} & Redback & \phn$16.7$ &0.70 & 2.00 & \nodata & \nodata & 23\\
\object[PSR J1311-3430]{J1311$-$3430} & BW & \phn$20.5$ &1.41 & 2.43 & \nodata & \nodata & 24\\
\object[PSR J1628-3205]{J1628$-$3205} & Redback & \phn42.1& 1.25 &1.22 & \nodata & \nodata & 25\\
\object[PSR J1810+1744]{J1810+1744} & Redback & \phn$39.7$ & 2.00 & 2.36 & \nodata& \nodata & 26\\
\object[PSR J1816+4510]{J1816+4510} & He WD & \phn$38.9$ & 2.42 & 4.36 & $4.5\pm1.7$& Photometry & 27\\
\object[PSR J1957+2516]{J1957+2516} & Redback & \phn$44.1$ &3.07 &2.66&\nodata & \nodata & 28\\
\object[PSR J2215+5135]{J2215+5135} & Redback & \phn$69.2$ & 3.01 & 2.78 & \nodata & \nodata & 27\\
\enddata
\tablerefs{(1) \citet{rsa+14}; (2) \citet{rhc+16}; (3) \citet{dvtb08}; (4) \citet{dcl+16}; (5) \citet{dab+12}; (6) \citet{rrb+15}; (7) \citet{mjds+18}; (8) \citet{nrh+11}; (9) \citet{scc+15}; (10) \citet{crr+16}; (11) Kaiser et al. (in prep.); (12) \citet{cls+13}; (13) \citet{kkv+15}; (14) \citet{bkb+16}; (15) \citet{rs11}; (16) Ray et al.\ (in prep.); (17) \citet{khc+12}; (18) \citet{scg+04}; (19) \citet{afw+13}; (20) \citet{gsl+16}; (21) \citet{bjs+16}; (22) \citet{kkn+16}; (23) \citet{drc+16}; (24) \citet{rfs+12};
(25) \citet{lht14}; (26) \citet{bvkr+13}; (27) \citet{kbvk+13}; (28) \citet{sab+16}.}
\tablenotetext{a}{We identify systems with helium-core white dwarf (WD) companions, Be binaries, redback systems, black widow (BW) systems), systems with main sequence (MS) companions, and transitional systems that alternate between accretion and rotation-powered.  PSR~J0337+1715 is a member of a triple system with two helium-core WDs.}
\tablenotetext{b}{DM-derived distance using the NE2001 model \citep{cl02}.}
\tablenotetext{c}{DM-derived distance using the YMW16 model \citep{ymw17}.}
\tablenotetext{d}{Methods used for DM-independent distance estimates: photometry and modeling of companion; direct VLBI astrometric parallax; combination of orbital period derivative (OPD) with kinematic models \citep{bb96}; direct pulsar timing parallax; spectroscopic parallax; association with star cluster or galaxy; modeling of lightcurve assuming Roche lobe overflow (RLO).}
\tablenotetext{e}{Using YMW16 in the Magellanic Cloud mode.}
\end{deluxetable}

\begin{deluxetable}{ l h c D D D c c c c c }
\tablecaption{\gaia\ astrometric measurements and derived parameters for pulsar companions\label{tab:measured}\label{tab:derived}}
\tablehead{
\colhead{PSR} & \nocolhead{Gaia DR2 Source ID} & \colhead{G mag.} & \twocolhead{$\px$} & \twocolhead{$\mu_{\alpha}$} & \twocolhead{$\mu_{\delta}$} & \colhead{$d_{\px}$} & \colhead{$d_{\mathrm{LK}}$} & \colhead{$v_{\perp,\px}$} & \colhead{$v_{\perp,\mathrm{LK}}$} & \colhead{$v_{\perp,\mathrm{DGR}}$}\\
 & & & \twocolhead{(mas)} & \twocolhead{(mas\,yr$^{-1}$)} & \twocolhead{(mas\,yr$^{-1}$)} & \colhead{(kpc)} & \colhead{(kpc)} & \colhead{(km\,s$^{-1}$)} & \colhead{(km\,s$^{-1})$} & \colhead{(km\,s$^{-1}$)}
}
\decimals
\startdata
\multicolumn{8}{l}{Pulsars with significant \gaia\ parallax measurements}\\
\cline{1-8}
J0337+1715 & 44308738051547264 & $18.08$ & $0.73(24)$ & $4.81(49)$ & $-4.42(42)$ & $1.11_{-0.12}^{+0.82}$ & $1.34_{-0.17}^{+1.07}$ & $34_{-4}^{+26}$ & $46_{-6}^{+44}$ & $48_{-6}^{+44}$\\
J0437$-$4715 & 4789864076732331648 & $20.41$ & $8.33(67)$ & $122.9(11)$ & $-71.2(16)$ & $0.118_{-0.007}^{+0.012}$ & $0.121_{-0.008}^{+0.013}$ & $79.5_{-5.1}^{+8.2}$ & $82.1_{-5.3}^{+9.0}$ & $78.8_{-5.4}^{+9.0}$\\
J1012+5307 & 851610861391010944 & $19.63$ & $1.33(41)$ & $2.98(52)$ & $-26.94(63)$ & $0.64_{-0.07}^{+0.42}$ & $0.79_{-0.09}^{+0.73}$ & $82_{-9}^{+55}$ & $113_{-12}^{+133}$ & $113_{-12}^{+132}$\\
J1023+0038 & 3831382647922429952 & $16.27$ & $0.73(14)$ & $4.75(13)$ & $-17.35(13)$ & $1.24_{-0.13}^{+0.39}$ & $1.32_{-0.14}^{+0.43}$ & $106_{-11}^{+33}$ & $116_{-13}^{+42}$ & $116_{-14}^{+46}$\\
J1227$-$4853 & 6128369984328414336 & $18.08$ & $0.62(16)$ & $-18.73(20)$ & $7.39(11)$ & $1.37_{-0.15}^{+0.69}$ & $1.70_{-0.17}^{+1.72}$ & $131_{-15}^{+66}$ & $180_{-15}^{+257}$ & $144_{-12}^{+187}$\\
B1259$-$63 & 5862299960127967488 & $9.63$ & $0.418(30)$ & $-6.986(43)$ & $-0.416(44)$ & $2.21_{-0.12}^{+0.19}$ & $2.26_{-0.13}^{+0.20}$ & $73.5_{-4.1}^{+6.1}$ & $75.4_{-4.3}^{+6.7}$ & $36.8_{-1.1}^{+1.3}$\\
J1417$-$4402 & 6096705840454620800 & $15.79$ & $0.221(71)$ & $-4.70(10)$ & $-5.10(87)$ & $3.50_{-0.38}^{+2.09}$ & $4.06_{-0.52}^{+2.24}$ & $115_{-13}^{+69}$ & $144_{-20}^{+86}$ & $118_{-15}^{+33}$\\
J1431$-$4715 & 6098156298150016768 & $17.75$ & $0.64(16)$ & $-12.01(33)$ & $-14.51(26)$ & $1.35_{-0.15}^{+0.69}$ & $1.70_{-0.15}^{+2.10}$ & $121_{-13}^{+59}$ & $169_{-10}^{+324}$ & $170_{-9}^{+263}$\\
J1723$-$2837 & 4059795674516044800 & $15.55$ & $1.077(54)$ & $-11.713(82)$ & $-23.990(62)$ & $0.90_{-0.04}^{+0.05}$ & $0.91_{-0.04}^{+0.05}$ & $114_{-5}^{+7}$ & $115_{-5}^{+7}$ & $124_{-5}^{+7}$\\
J2032+4127 & 2067835682818358400 & $11.36$ & $0.693(33)$ & $-2.991(48)$ & $-0.742(55)$ & $1.38_{-0.06}^{+0.07}$ & $1.39_{-0.06}^{+0.08}$ & $20.1_{-0.8}^{+1.1}$ & $20.4_{-0.9}^{+1.2}$ & $25.7_{-1.0}^{+1.2}$\\
J2129$-$0429 & 2672030065446134656 & $16.84$ & $0.424(88)$ & $12.34(15)$ & $10.19(14)$  & $2.06_{-0.21}^{+0.67}$ & $2.18_{-0.23}^{+0.72}$ & $157_{-16}^{+51}$ & $172_{-19}^{+61}$ & $191_{-23}^{+77}$\\
J2339$-$0533 & 2440660623886405504 & $18.97$ & $0.75(25)$ & $4.15(48)$ & $-10.31(30)$ & $1.08_{-0.12}^{+0.82}$ & $1.25_{-0.16}^{+0.84}$ & $57_{-6}^{+47}$ & $72_{-10}^{+56}$ & $64_{-9}^{+54}$\\
\hline
\multicolumn{8}{l}{Pulsars with low significance \gaia\ parallax measurements}\\
\cline{1-8}
J0045$-$7319 & 4685849525145183232 & $16.22$ & $0.040(60)$ & $0.31(11)$ & $-0.931(99)$ & \nodata & \nodata & \nodata & \nodata & \nodata\\
J0348+0432 & 3273288485744249344 & $20.64$ & $-2.0(10)$ & $3.1(20)$ & $-0.1(14)$ & \nodata & \nodata & \nodata & \nodata & \nodata\\
J1024$-$0719 & 3775277872387310208 & $19.18$ & $0.53(42)$ & $-35.52(64)$ & $-47.93(65)$ & \nodata & \nodata & \nodata & \nodata & \nodata\\
J1048+2339 & 3990037124929068032 & $19.65$ & $0.96(80)$ & $-16.3(10)$ & $-11.7(12)$ & \nodata & \nodata & \nodata & \nodata & \nodata \\
J1311$-$3430\tablenotemark{a} & 6179115508262195200
 & $20.53$ & \multicolumn{2}{c}{\nodata} & \multicolumn{2}{c}{\nodata} & \multicolumn{2}{c}{\nodata} & \nodata & \nodata & \nodata & \nodata & \nodata\\
J1628$-$3205 & 6025344817107454464 & $19.52$ & $1.20(56)$ & $-6.4(10)$ & $-19.81(82)$ & \nodata & \nodata & \nodata & \nodata & \nodata\\
J1810+1744 & 4526229058440076288 & $20.08$ & $1.05(69)$ & $6.4(17)$ & $-7.2(20)$ & \nodata & \nodata & \nodata & \nodata & \nodata\\
J1816+4510 & 2115337192179377792 & $18.22$ & $0.22(15)$ & $-0.17(29)$ & $-4.42(33)$ & \nodata & \nodata & \nodata & \nodata & \nodata\\
J1957+2516 & 1834595731470345472 & $20.30$ & $0.69(86)$ & $-5.7(11)$ & $-8.9(14)$ & \nodata & \nodata & \nodata & \nodata & \nodata\\
J2215+5135 & 2001168543319218048 & $19.24$ & $0.28(36)$ & $0.31(54)$ & $1.88(60)$ & \nodata & \nodata & \nodata & \nodata & \nodata\\
\enddata
\tablecomments{Quantities in parentheses represent the 1-$\sigma$ uncertainties reported by \gaia.  Along with astrometric parameters, we give the \gaia\ G-band mean magnitude of the associated source. Pulsars with detected companions (\S~\ref{sec:id}) but parallax measurements with errors greater than 40\% of their values are shown below the line. For each pulsar, $d_{\px}$ is the distance estimate calculated assuming a uniform prior on parallax, and may be considered a maximum-likelihood estimate, while $d_{\mathrm{LK}}$ is the Lutz-Kelker corrected distance, calculated using the volumetric prior described in Section~\ref{sec:lk}. Both distance estimates take into account the zero-point offset of $-0.029$ mas given by \citet{lhb+18}. Velocities are calculated using the \gaia\ proper motions given here. The transverse velocity $v_{\perp,\px}$ of the pulsar is calculated using $d_{\px}$ as a distance estimate, while $v_{\perp,\mathrm{LK}}$ is calculated using $d_{\mathrm{LK}}$. The third velocity $v_{\perp,\mathrm{DGR}}$ is an estimate of the pulsar's velocity with respect to its standard of rest. It also uses $d_{\mathrm{LK}}$ as a distance estimate, and includes corrections due to differential Galactic rotation and the peculiar motion of the Sun. For all distances and velocities, we give the posterior mode as a point estimate, and the 16th and 84th percentiles of the posterior distribution as errors.
\tablenotetext{a}{The companion to PSR J1311$-$3430 is detected in \gaia\ DR2, but only the two-parameter astrometric solution (position on the sky) is reported, because the five-parameter solution (including parallax and proper motion) failed the acceptance criteria detailed in~\citet{lhb+18}. A full solution is expected in future data releases, and so it is included here for the sake of completeness. }}
\end{deluxetable}

Of the pulsars with companions seen by \gaia,\ five also have pulsar timing or VLBI parallaxes, as indicated in Table~\ref{tab:other}. With the exception of PSR~J0437$-$4715 (see below), all of these distance estimates are consistent with our \gaia\ DR2 results. Of these, PSR~J1024$-$0719 does not have a significant parallax measurement in DR2. In the cases of PSRs~J0437$-$4715 and J1012+5307, the most precise distance estimates have been achieved through orbital period derivative modeling \citep{rhc+16,dcl+16}, and are significantly more precise than the \gaia\ distances. For PSR~J1023+0038 the VLBI parallax is more precise than the \gaia\ value, while the reverse is true for PSR~B1259$-$63.

In Figure~\ref{fig:compare2} we compare the \gaia\ distances with the DM distances.  We see that both DM models do reasonably well in comparison with \gaia, with YMW16 showing somewhat better agreement for the sample of objects considered here. 
An assumed 25\% uncertainty gets the majority of points consistent between the astrometric and DM distances. However, the distances predicted by both DM models appear to be somewhat underestimated compared to those measured by \gaia---the median ratio of DM distance to \gaia\ distance is 0.76 for NE2001 and 0.90 for YMW16.
This agrees roughly with the findings of \citet{gmc08} and \citet{r11}, who claimed that for pulsars out of the Galactic plane (which is true for most recycled pulsars), distances predicted by the NE2001 model were underestimated. 

We compare the proper motions from \gaia\ and the literature in Figure~\ref{fig:pm}.  Again the agreement for most sources is reasonably good, with the sources with the most precise prior measurements (PSRs~J0437$-$4715, J1012+5307, J1023+0038) agreeing very well.

\subsection{Notes On Individual Objects}
\begin{description}
\item[J0337+1715] This pulsar is in a hierarchical triple system with two white dwarfs \citep{rsa+14}. The orbital period of the outer white dwarf companion is 327 days, which is comparable to the 1-year parallactic period and not too much smaller than the 22-month data span of \gaia\ DR2. As a result, the orbital motion of the inner white dwarf (which is the optically dominant one) with the outer white dwarf may have caused systematic errors in the proper motion and (especially) parallax measurements.  However, we find reasonable agreement with the distance inferred from white dwarf photometry \citep{kvkk+14} and from VLBI astrometry (Deller et al., in prep.).

\item[J0437$-$4715] This is the closest system in our sample by a factor of about 5 (and one of the closest known neutron stars). The most precise distance measurement comes from orbital period derivative modeling \citep{rhc+16}. The \gaia\ distance is only about 77\% of this value, and the results are formally inconsistent.  The inconsistency may be due to unmodeled binary reflex motion, as the companion should trace out an ellipse with size $\approx 0.3\,$mas (based on the masses from \citealt{vbvs+08}) every 5.5\,d, comparable to the size of the parallax uncertainty.  Most other systems (with the exceptions of the Be binaries discussed below, the outer companion of PSR~J0377+1715, and PSR~J1024$-$0719; see \citealt{bjs+16,kkn+16}) have much smaller projected orbital separations.  The discrepancy may be resolved with future \gaia\ data releases when the individual astrometric measurements are released and binary orbits are included in the fits.

\item[B1259$-$63] The system orbital period is 1237 days, which less than twice the 22-month span of the data used to derive the \gaia\ astrometric parameters, and may have caused residual systematic errors in those parameters.

\item[J1417$-$4402] Here we find that the \gaia\ distance constraint is much more consistent with the distance estimate from assuming that the companion fills its Roche lobe \citep[4.4\,kpc,][]{scc+15,crr+16} than the dispersion-measure distance of 1.6\,kpc.

\item[J1816+4510] The \gaia\ proper motion $\mu_\alpha$ disagrees significantly with the one from pulsar timing, a discrepancy to be investigated further.  

\item[J2032+4127] This is a long-period ($46\pm2$\,yr) binary system with a wide orbit.  Reflex motion of the companion may have affected the \gaia\ astrometric fit, especially for proper motion.

\item[J2215+5135] While there is no significant parallax for this source, the \gaia\ proper motion (magnitude $1.9\pm0.8\,{\rm mas\,yr}^{-1}$) is much smaller than the value listed in \citet[][magnitude $189\pm23\,{\rm mas\,yr}^{-1}$]{aaa+13}.  We believe the latter to be an error: it would also imply an unlikely transverse velocity of $2700\,{\rm km\,s}^{-1}$ at the DM distance.
\end{description}

\section{Discussion}
\label{sec:discuss}
As discussed above, we focus on binary systems since they are, with the exception of the Crab pulsar, the only pulsar systems with optical emission detectable by \gaia.  We find that the systems we have identified fall into 3 broad categories: recycled pulsars with white dwarf companions; recycled pulsars with non-degenerate companions (so-called ``black widows'' and  ``redbacks''; \citealt{r13}); and young pulsars with massive B star companions.

Both of the first two classes form from similar progenitors, where millisecond pulsars (MSPs) are produced in binary systems when old NS are ``recycled'' by accreting matter from the companion star. After recycling, the evolution of such systems can take a number of paths, with some diverging and leaving white dwarf companions and some converging and gradually destroying their companions through accretion and ablation \citep{t11,ccth13}.  Some of the converging systems are visible as redbacks, typically with distorted main-sequence-like companions of a few tenths of a solar mass, or black widows with companions of a few hundredths of a solar mass \citep{r13}. Some of the systems that we identified are still in the process of transitioning from accreting to rotation-powered \citep[e.g., the transitional millisecond pulsar J1023+0038;][]{asr+09}, but resemble redbacks when not accreting.  For the white dwarf systems, \gaia\ detections are biased in favor of the lower-mass (and hence larger) companions, especially as many of them are helium-core white dwarfs that can have hydrogen shell burning (stable or sporadic) which keeps them hotter and brighter for Gyr \citep[e.g.,][]{ab98,dbsh99,imt+16}.

The young pulsars such as PSR~B1259$-$63 follow a different evolutionary path, with wide orbits around massive (and luminous) stars.  However, all of these systems share characteristics that made them suitable for \gaia\ detections: they are all both bright and nearby.

Our sample above contains a number of objects of individual interest.  Some are used as laboratories for tests of alternative theories of gravity (PSR~J0348+0432, \citealt{afw+13}; PSR~J0337+1715, \citealt{rsa+14}), while others (PSR~J1023+0038, \citealt{sdg+18}) help test theories of accretion,   and the Be systems are testbeds for particle acceleration (PSR~B1259$-$63, \citealt{d13}).

The ensemble of distances to the sample of recycled pulsars is also of particular interest.  A major effort is underway to regularly time MSPs in pulsar timing arrays (PTAs) to detect low-frequency (nHz) gravitational waves \citep[e.g.,][]{bjpw13,d79,l15}. Precise measurements of the distances to MSPs can significantly improve the PTA sensitivity  when searching for coherent signals from individual sources \citep{mzh+16}. The most relevant pulsars for PTA purposes are the most stable, particularly fully-recycled millisecond pulsars which typically have low-mass white dwarf companions. Our sample includes four such systems, as well as the partially-recycled pulsar PSR~J0348+0432. \gaia\ provides the first independent distance estimate for PSR~J1431$-$4715, while for PSR~J1012+5307, it provides a distance estimate with precision comparable to that achievable with pulsar timing. However, the \gaia\ parallax distance estimates cannot compete with the precision achieved with VLBI for PSR~J0437$-$4715, and the \gaia\ parallax measurement for the distant PSR~J1816+4510 will likely remain insignificant until the end of \gaia's nominal mission (see below).

\gaia\ DR2 is based on a $T=22$\,month  data span. The final \gaia\ data release, planned for 2022, will be based on data collected during \gaia's\ entire five-year nominal mission. Since errors in parallax measurements scale as $T^{-1/2}$ and errors in proper motion scale as $T^{-3/2}$, the additional 38 months of data should allow the errors in parallax to be reduced to approximately 60\% of their current values, while the errors in proper motion should decrease by a factor of approximately 4.

While \gaia\ DR2 is essentially complete for sources with magnitude $G<17$~\citep{gaia-dr2}, many pulsar companions are fainter than that limit, and it is likely that at least some new pulsar companions will be present in the final data release, which is expected to be complete to magnitude $G\approx 20.7$~\citep{gaia}. As mentioned previously, redbacks and other pulsars with main sequence-like companions present the best targets for \emph{Gaia}, but the 7 redback systems and 3 other pulsars with main sequence-like companions given in Table~\ref{tab:other} already represent the majority of known examples of such systems. It is most likely that new pulsar companions detected in forthcoming \gaia\ data releases will represent both redbacks and millisecond pulsars with white dwarf companions, several of which also appear in the DR2 data discussed here.

In \gaia\ DR2, all sources are modeled as single stars~\citep{gaia-dr2}. This raises the possibility that parallax and proper motion measurements for objects in binaries, like those considered here, may be corrupted by unmodeled orbital motion. A possible indicator of this corruption is the \gaia\ excess noise parameter \citep[\texttt{astrometric\_excess\_noise} in the \gaia\ Archive]{lhb+18}, which is a measure of the size of effects not accounted for by the single-star model. Most companions to redback pulsars, and some Be stars, also show significant flux variability, but this is unlikely to have a systematic effect on the \gaia\ parallax estimates, since the period of the flux variations is not a simple fraction of Earth's orbital period. For the majority of the sources in Table~\ref{tab:measured}, corruption by orbital motion is unlikely to be an issue, as their orbital periods are small compared to the 22-month observing time of \gaia\ DR2, and any orbital motion will average out. For the nearest pulsar in our sample, PSR~J0437$-$4715, the orbit of the white dwarf companion spans $0.3$ mas on the sky, which is about half the uncertainty in its parallax. However, its orbital period is 5.74 days, meaning that it completes approximately 64 orbits in a year, and 116 in the 22-month \gaia\ DR2 observing span, so the contribution of this orbital motion to the parallax and proper motion measurements should largely average out. Additionally, \gaia\ reports no excess noise for the source associated with this pulsar, which is perhaps surprising considering that its parallax gives a distance which is somewhat inconsistent with the highest-precision value derived from orbital period derivative modeling.

On the other hand, PSR~B1259$-$63 has an orbital period of 1237 days, and the triple system PSR~J0337+1715 has an outer orbital period of 327 days, both of which are comparable to the observing time, and so the \gaia\ parallaxes and proper motions of these pulsars may be affected by a systematic error due to orbital motion.  That appears to be evident in the quality of the \gaia\ astrometric fit and the parallax uncertainty compared with what \gaia\ predicts for a source of that $G$ magnitude\footnote{Using the formula from \url{https://www.cosmos.esa.int/web/gaia/science-performance} scaled by 1.5 to account for the limited duration of DR2.}.  For PSR~J0337+1715 the reduced $\chi^2$ of the astrometric fit is 1.8, on the high end for our sources, and the measured parallax uncertainty of 0.25\,mas agrees very well with the predicted uncertainty of 0.21\,mas.  For PSR~B1259$-$63 the fit reduced $\chi^2$ is even higher at 2.0, and the measured parallax uncertainty of 0.03\,mas exceeds the predicted uncertainty of 0.01\,mas.  The only other source with a high reduced $\chi^2$ and a large ratio of measured to predicted parallax uncertainty is PSR~B2032+4127: like PSR~B1259$-$63, this is a Be binary with a bright ($G=11.4$) companion. For PSR~J0337+1715, the excess noise reported by \gaia\ is approximately 0.3 mas, while for PSR~B1259$-$63, no excess noise is reported. The only other pulsar with a significant parallax and nonzero excess noise parameter is PSR~J1431$-$4715, for which the excess noise is again around 0.3 mas. So while there may be some corruption of the astrometric fits due to unmodeled orbital motion, the bigger effect may just be an overestimate of the attainable precision for the brightest sources.  The full \gaia\ data-sets with longer timespan,  better calibration, and binary model fits, will help to resolve the question.  Ultimately, the \gaia\ data represent a small but potentially important contribution to the pulsar distance scale, and will be incorporated in future versions of the pulsar population analysis and electron density models.

\acknowledgments
We thank C.~Mingarelli for valuable discussions regarding pulsar cross-matching and the parallax zero-point offset, and W.~Vlemmings for helpful comments. Additionally, we thank the anonymous referee for providing useful feedback.
RJJ, DLK, SC, JMC are members of the NANOGrav Physics Frontiers Center, which is supported by the National Science Foundation award 1430284. ATD is the recipient  of  an Australian  Research  Council  Future  Fellowship (FT150100415).

This work has made use of data from the European Space Agency (ESA)
mission {\it Gaia} (\url{https://www.cosmos.esa.int/gaia}), processed by
the {\it Gaia} Data Processing and Analysis Consortium (DPAC,
\url{https://www.cosmos.esa.int/web/gaia/dpac/consortium}). Funding
for the DPAC has been provided by national institutions, in particular
the institutions participating in the {\it Gaia} Multilateral Agreement.

The Pan-STARRS1 Surveys (PS1) and the PS1 public science archive have been made possible through contributions by the Institute for Astronomy, the University of Hawaii, the Pan-STARRS Project Office, the Max-Planck Society and its participating institutes, the Max Planck Institute for Astronomy, Heidelberg and the Max Planck Institute for Extraterrestrial Physics, Garching, The Johns Hopkins University, Durham University, the University of Edinburgh, the Queen's University Belfast, the Harvard-Smithsonian Center for Astrophysics, the Las Cumbres Observatory Global Telescope Network Incorporated, the National Central University of Taiwan, the Space Telescope Science Institute, the National Aeronautics and Space Administration under Grant No. NNX08AR22G issued through the Planetary Science Division of the NASA Science Mission Directorate, the National Science Foundation Grant No.\ AST-1238877, the University of Maryland, Eotvos Lorand University (ELTE), the Los Alamos National Laboratory, and the Gordon and Betty Moore Foundation.  The national facility capability for SkyMapper has been funded through ARC LIEF grant LE130100104 from the Australian Research Council.
SkyMapper is owned and operated by The Australian National University's Research School of Astronomy and Astrophysics. 

\facilities{Gaia}
\software{Astropy \citep{art+13}, Astroquery \citep{gsp+18}}

\clearpage
\bibliographystyle{aasjournal}

\end{document}